# Proposal for a

# High Precision Tensor Processing Unit

## RNS TPU

**Version 1.01**

With
Relevance of Number System Design
And its impact on
General Purpose Computation

Original release V1.0: June 8th, 2017
Updated release V1.01: June 9th, 2017

Whitepaper

By

Eric B. Olsen

Digital System Research, Inc.

## Abstract

This whitepaper proposes the design and adoption of a new generation of Tensor Processing Unit (TPU) which has the performance of Google's TPU, yet performs operations on wide precision data. The new generation TPU is made possible by implementing arithmetic circuits which compute using a new general purpose, fractional arithmetic based on the residue number system (RNS).

# Table of Contents

## Introduction

Google's Tensor Processing Unit (TPU) has recently gained attention as a new and novel approach to increasing the efficiency and speed of neural network processing. According to Google, the TPU can compute neural networks up to 30x faster and up to 80x more power efficient than CPU's or GPU's performing similar applications[6]. The TPU excels because its hardware processing flow is specifically adapted to the inference problem it solves. It has been shown this inference task can be programmed to operate using 8 bit data; by optimizing arithmetic circuits to operate specifically on this narrow data width, the TPU has been able to operate faster by placing more multipliers in parallel (on a given die size), thereby breaking barriers set by traditional processors, such as CPUs and GPUs.[5]

It is the goal of this paper to disclose that a new form of general purpose arithmetic, having recently emerged, is an ideal fit to power a new generation of TPU. The new generation of TPU will have the advantage of supporting wide precision arithmetic, yet retain the awesome performance and energy efficiencies of the original Google TPU.

## TPU Architecture

In simple terms, the TPU is a hardware based matrix multiplier supporting a 256x256 matrix of 8 bit hardware multipliers. (See block diagram of Figure 1 redrawn from reference [5] with control and data rate detail removed.) One innovative feature is the systolic shifting circuitry allowing data to flow into the hardware matrix thus providing 65,536 multiplies every second [6]. A more subtle feature is a result of tuning the neural network calculations to use 8 bit data values. Specifically, multiplication is performed on two 8 bit integer values, with the resulting 16 bit product summed using a 32 bit accumulator, thereby forming the required dot product (product summation). By summing 16 bit values, normalization is delayed, and a summation value with a larger dynamic range (say up to 24 bits) is then presented to the activation step where the data is processed using a ReLU, Sigmoid or other function . After the activation step, the data is normalized to restore it to the original 8 bit data format. The resulting normalized data can be stored and re-used by the matrix multiplier depending on the NN algorithm requirements.

In the case of the TPU, delaying normalization is fast and convenient, since 8x8 bit multipliers and 32 bit accumulators are fast, common components in ASIC libraries. This all works as the inference phase of the NN's used at Google will operate with negligible error. But this may not be the case for the training phases of Google's NN algorithms. For this, wider precision data is used, and therefore Google will process NN training phases using GPU based solutions. While there are other reported successes at quantizing data for NN's to improve arithmetic efficiency [5][11], there are certainly algorithms which fail to operate using quantized data, even in the case of the quantization of 32 bit fixed point to 16 bit fixed point [12]. In addition, other NN algorithms now and in the future may not operate sufficiently with 8 bit data.

If wish to support a higher precision arithmetic, what happens if we increase the data width of the TPU architecture?

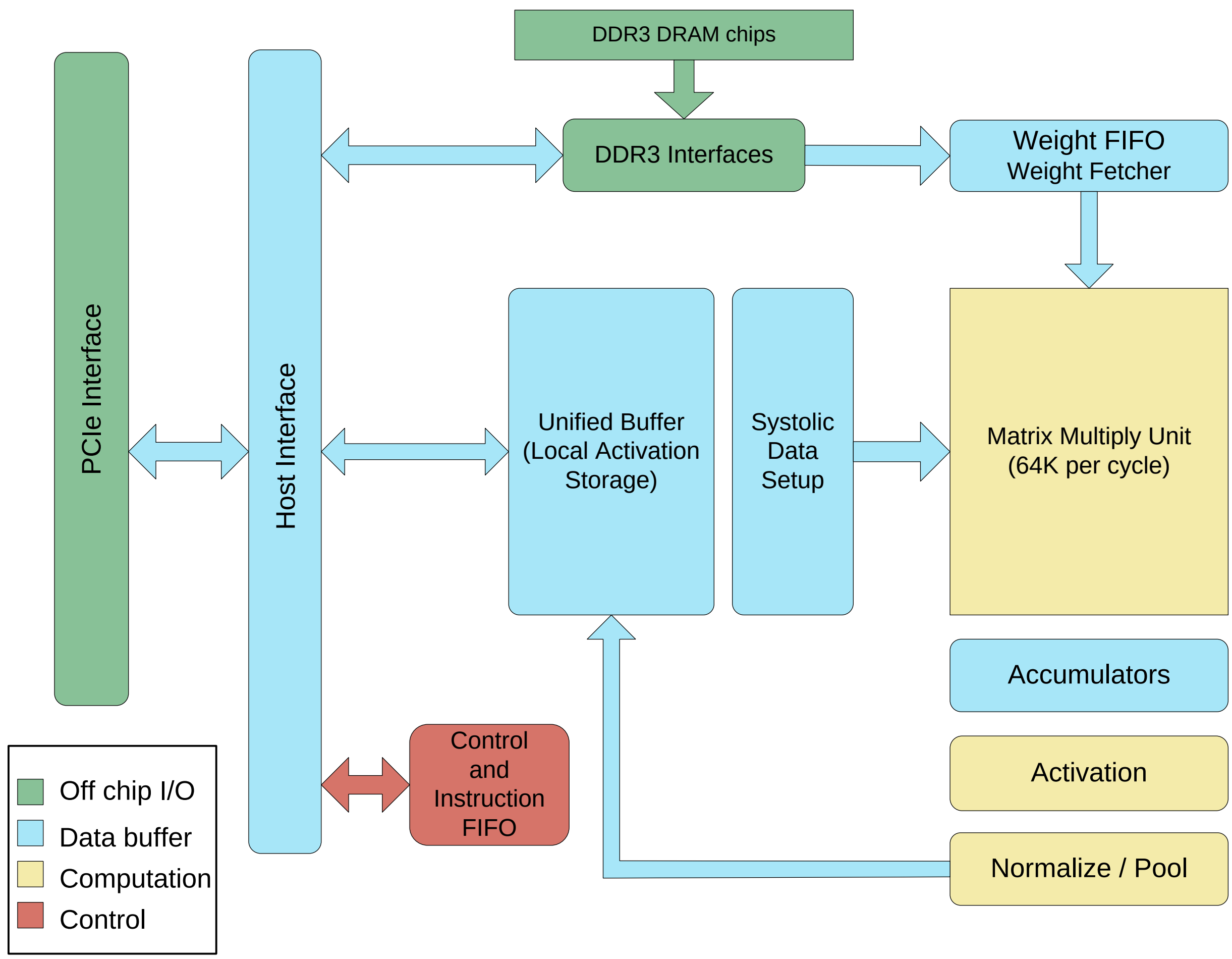

TPU Block diagram re-drawn from Figure 1 of reference[6]
with control detail and bandwidth rates removed

Figure 1

## Increasing data width of Google TPU

If we propose for a moment to increase the data width of the TPU multipliers, and for the purpose of argument, we stick with a 256x256 array, things start to grow out of bounds quickly. If we increase the 8 bit multiplier to a 16 bit multiplier, our product results are 32 bits wide, and accumulations are 40 bits wide. If we adopt a 32 bit multiplier, we obtain 64 bit product results, and 72 bit wide accumulations. *We can deduce there is a tipping point where the process of delaying normalization is counter-productive because carry delay becomes problematic.* We see other associated issues, such as the rapid increase in the area of multipliers, and the increasing size of data buses, which significantly complicates the design of the systolic circuitry which feeds the hardware multiplier matrix. Larger buses and larger multipliers mean longer signal paths, and this introduces increased power consumption and additional delays.

Therefore, we cannot increase the data width of the Google TPU and expect to keep the same speed and efficiency.

Google and other top tier cloud companies are driven to constantly seek improvement in processing efficiency and speed in order to satisfy the ever increasing demand on its data servers. With the real possibility that Moore's law slows, what can companies like Google do to improve speed and efficiency?

## Improving hardware architecture (may not be enough)

Designers often look toward architectural improvements to gain speed in their designs. But this subject is well established, as many techniques have been developed to improve speed through improved hardware architecture. In fact, CPU's, GPU's and FPGA's often battle in terms of offering better performance; this better performance is primarily derived from architectural efficiencies of the underlying arithmetic hardware. Therefore, all things being equal, it is evident that architecture alone is not providing *giant* leaps in performance, albeit, with the exception of the novel TPU by Google. But as shown above, simply making the TPU multipliers and data paths wider is not a beneficial architectural improvement if we wish to process wide data at the same rates, and same efficiencies, compared to the TPU.

Painted into a corner, designers must seek radical new approaches if they are to succeed in significantly improving the speed and efficiency of hardware and if they hope to minimize the impact of a slowing Moore's law. Google's TPU design team is hereby commended!

## Looking to the future … and to the past

Obviously, in the field of digital computer design, designers look to the future to improve speed and efficiency. Well known is the fact that shrinking process technology produce faster and denser circuitry, which drives Moore's law, and generally we all share this advantage. More exotic solutions, such as Quantum computing and Optical computing are futuristic, but are barred from general purpose computations by design, practicality and the state of the art of these technologies. For example, commercially available Quantum computers are not digital (as of this writing), nor are they deterministic, and therefore cannot process general purpose arithmetic like a digital computer.

When painted into a corner, some designers look back; by scouring discarded ideas from the past, some hope a resurrected idea might point the way to improved performance and efficiency. Interestingly, the TPU uses a systolic approach to its matrix data flow; systolic methods have historical significance.

Another popular idea from the 50's and 60's is to encode data into a format that eliminates the need for carry. This approach was motivated by a need to reduce the size of multiplier circuitry, and also to reduce delays of carry propagation. As process technology shrunk, solutions involving carry free number systems fell out of commercial consideration. But could these abandoned techniques point a way to improved performance and efficiency in today's modern binary world?

## Revisiting carry-free number systems

There are a number of unique binary encodings that allow at least some arithmetic operations to be performed without carry. One such encoding is called the "residue number system" or RNS for short. The earliest indication of the residue number system was stated in a poem by the Chinese mathematician Sun-Tsu. An early use of RNS in computer arithmetic was by Valach in 1958 [1]. Another early pioneer, Harvey Garner, made significant contributions to the field in 1959.

Probably the most significant amount of information regarding RNS was compiled in research labs during the 1960's. Fearing the information might be lost, researchers Szabo and Tanaka decided to compile this information in a publication entitled "Residue Arithmetic and its Applications to Computer Technology" [1]. This was an enlightened act, as the hope of RNS being used in commercial applications quickly faded. However, in the 70's, 80's, 90's and beyond, a decent body of excellent research in RNS continued at the university level [4][13][15] and in some commercial research labs. For example, significant successes in implementing FIR filters have been implemented using basic RNS arithmetic [4][15]. But these are R&D prototypes. This author is not aware of any significant use of RNS processing found in commercially competitive hardware. So in the author's opinion, RNS has *not* played a significant role in modern computer systems.

To make a long story short, the RNS system failed to become useful for a myriad of reasons. While RNS addition and multiplication is indeed carry free, other problems thwarted this advantage; often cited are: 1) conversion of binary to RNS, 2) conversion of RNS to binary, 3) base extension, 4) comparison, 5) division, and many other functions. Perhaps contributing as much to the failure of RNS was the paradigm of the day. That is, in the design of an RNS based *binary* multiplier, a common approach was to first convert binary data to RNS format, then multiply the RNS data, and finally convert the RNS product result back to binary. Figure 2 illustrates the failed approach. The problem is that delays and inefficiencies of forward and reverse conversion out-weigh any advantage gained by the carry free RNS multiplication.

Another issue is more subtle but is fundamental. Figure 2 shows a prior art approach, whereas existing binary operations prevail in their entirety except for a relatively isolated attempt to replace integer multiplication and addition operations with RNS operations. In this paradigm, a general purpose ALU would therefore require the result of the multiplication be converted back to binary as soon as possible to allow further processing in binary format. Again, the "sandwiching" of two layers of conversion for each RNS multiply and accumulate is no faster than simply performing a binary MAC operation directly.

To summarize early efforts, RNS based arithmetic, while exhibiting amazing carry-free properties, had too many problems to replace binary multipliers and adders during the computer revolution of the 60's. In the 1970's, as the revolution of circuit miniaturization was in full swing, the ability to integrate advanced carry and multiplier circuitry directly on-chip changed the design landscape, so the use of RNS fell out of practical consideration and was relegated to an advanced research topic.

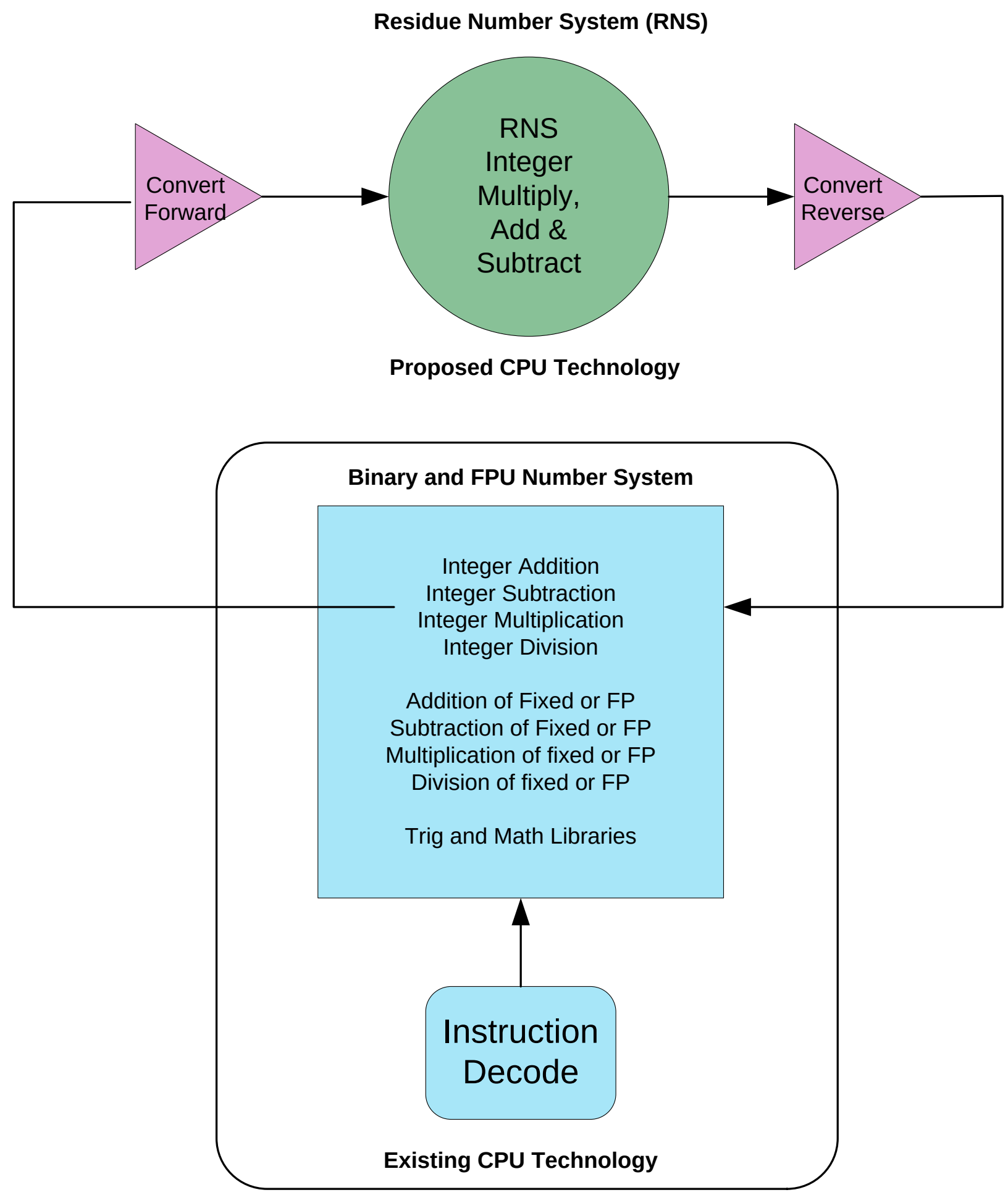

Figure 2.

## The relevance of number systems

It is relevant to this proposal and to this story that we consider a few aspects of the historical development of number systems. Please be nice, we can't cite everybody!

The decimal number system, for example, was used by Egyptians 4000 years ago but was not positional. More than a thousand years later, Babylonians helped invent the notion of a positional system with its base 60 Sexagesimal number system. The invention of the positional *decimal* number system with number zero has origins in India about 500 AD. Similarly, South American Olmecs and Mayans use the number zero at about this time. In the $7^{th}$ century, the noted Persian mathematician al-Khwarizmi was instrumental in many advancements and clarifications of the decimal number system including an explanation of the use of zero. Many other contributors exist throughout the world not mentioned. Now nearly 600 years old, the Hindu-Arabic decimal number system was introduced in Europe in the $11^{th}$century. But amazingly, all of these variations of the decimal number system were primarily for *integers only*. While fractions existed, they were mainly considered as ratios of two (decimal) integers.

In fact, it wasn't until 1585 that Simon Stevin published his famous pamphlet "One Tenth", which described an orderly approach to using the decimal number system to represent fractional quantities. Until that time, mathematicians wrote fractions primarily as a ratio or summation of specific ratios, so that the corresponding arithmetic of fractions using this notation could be quite difficult for most. Imagine adding a summation of ratios [1/100 + 1/50 + 1/5] with another summation of ratios [1/25 + 1/10 + ½], yet you don't have a fractional decimal system to work with!

Stevin's approach to decimal fractional representation and arithmetic is essentially taught in every grade school throughout the world today. While Stevin himself admits to not having invented fractions, Stevin is credited with formalizing and defining fractional decimal notation and arithmetic in an orderly fashion. His contribution might seem trivial. However, Stevin made fractional arithmetic (by hand) *practical*, therefore, *Stevin made general purpose computation practical,* and therefore *possible*!

History shows it took centuries to conceive of and adopt a practical fractional number system. Furthermore, the decimal number system spent considerable time as an "integer only" number system during its long development. But once Stevin defined decimal fractions, it might not appear a difficult stretch to formalize this sort of number system into a common category, what we now call *fixed radix*. But amazingly, the binary number system, indeed a fixed radix system, was not published until 60 years later by Gottfried Leibniz. But when Leibniz invented his Step Reckoner calculator to perform arithmetic, he designed a machine encoded with decimal [7], presumably since humans think this way!

It's somewhat remarkable that Babbage didn't see an electrical solution for his computers, being that he understood binary numbers, and that electric relay technology in the form of telegraph equipment was wide spread. After all, Babbage was a remarkably talented thinker and inventor. But Babbage was obsessed with mechanical computer design; his second machine was a programmable digital computer designed to solve polynomial equations. Arithmetic was processed using many mechanical digits "encoded" with the decimal number system. A programmable computer designed to solve certain arithmetic had been *envisioned*, but it was not general purpose.

It took another 100 years or so to get the ball rolling again, this time in the form of electro-mechanical and electronic computers. Initially, computer design was aimed at solving specific classes of calculations. Then, *rather quickly, general purpose operation was sought*. Consider that some early electronic computers did not adopt binary, but adopted the decimal number system, once again. The *unilateral* adoption of binary for machine computation took decades. But it did occur. Certainly, binary was the best number system for machine computation. And certainly, human understanding of number systems was getting better, right? Well, *maybe not*!

## The amazing mischaracterization of RNS

In most all texts discussing residue numbers, we are introduced to RNS as being an "*integer only*" number system [1][2][3]. Sound familiar? Once again we see a similarly narrow view of a number system being propagated, despite the fact that many very intelligent people have studied it. But perhaps it's not so amazing; the RNS fractional format had not been developed, and so it is essentially ignored, relegated to the impractical, or even relegated to the impossible. Perhaps missing was any motivation to push RNS this far forward.

Let's be clear, the RNS number system is mildly cryptic, and in no way resembles or operates like a fixed radix number system; yet recently, DSR has found that the underlying mathematics for encoding a fractional number in RNS is nearly identical to that of a fixed radix number system. While the underlying encoding is similar, digit and word operations required to carry out fractional arithmetic are quite different.

In 2012, the USPTO published patent application US20130311532 by this author [14]. The application discloses a practical definition for fractional representation in RNS, and also discloses the algorithms and procedures for operating on these fractional RNS formats. Also disclosed are important apparatus, such as hardware to support fractional RNS multiplication and fractional RNS division; furthermore, apparatus is disclosed to support fractional binary to fractional RNS conversion, and fractional RNS to fractional binary conversion. Prior art RNS operations are included as well, such as basic addition and subtraction, integer multiplication, base extension, comparison, and the encoding of negative numbers. Other essentially unsolved operations, such as arbitrary integer division, are disclosed and help build a complete set of operations to enable general purpose processing in RNS.

Moreover, US20130311532 argues for a coherent strategy for sustained, iterative, and repetitive calculation entirely in residue number format. To accomplish this, we must consider the objective: promoting *a new form of general purpose arithmetic*, i.e., a new form of general purpose machine computation. Considering the historical perspective, the task of upgrading an "integer only" RNS to a system suitable for performing general purpose arithmetic is apparently large, as there are plenty of lessons and new perspectives to be learned in that process. From the perspective of classical RNS, the task includes addressing and solving many known problems of RNS arithmetic. And finally, one must design the mathematics, the digit operations, and the hardware design for a complete set of arithmetic functions. This complete set of arithmetic hardware must operate in a *coherent* manner, such that digit and word manipulations within and between sub-modules work together to support sustained, general purpose calculation in RNS format.

## Development of the Rez-9

In 2012, this author embarked on the development of an RNS ALU prototype using an Altera FPGA. The first ALU design was designated Rez-1; it supported 18 residue digits, with each digit encoded using a 6 bit binary value. Table look-up was used to support most digit arithmetic operations. During the design process, much was learned about synthesizing modular arithmetic into hardware built for binary logic. Before the Rez-1 design was complete, it became evident that better techniques could be applied to empower the ALU design, thus the Rez-9 project was born.

In the summer of 2013, Digital System Research, Inc (DSR) accelerated the design and development of a second generation RNS ALU called Rez-9. Daniel Anderson, a graduate student at UNLV, joined the team and was tasked with implementing instruction execution and control circuitry previously modeled and implemented in software. Working together, Olsen and Anderson completed a hardware prototype of the Rez-9 in December of 2014. One day before the presentation of the Rez-9 as part of Anderson's thesis, the Rez-9 successfully executed a Mandelbrot plotting operation, which demonstrated the first sustained, iterative, *fractional* RNS processing in hardware. (software simulations of RNS Mandelbrot calculations had been developed several years earlier by Olsen). The next day, the Rez-9 prototype was publicly demonstrated for the first time by Daniel Anderson during his thesis presentation at UNLV [10]. Thus, *RNS based general purpose processing is defined and possible, but will it be practical*?

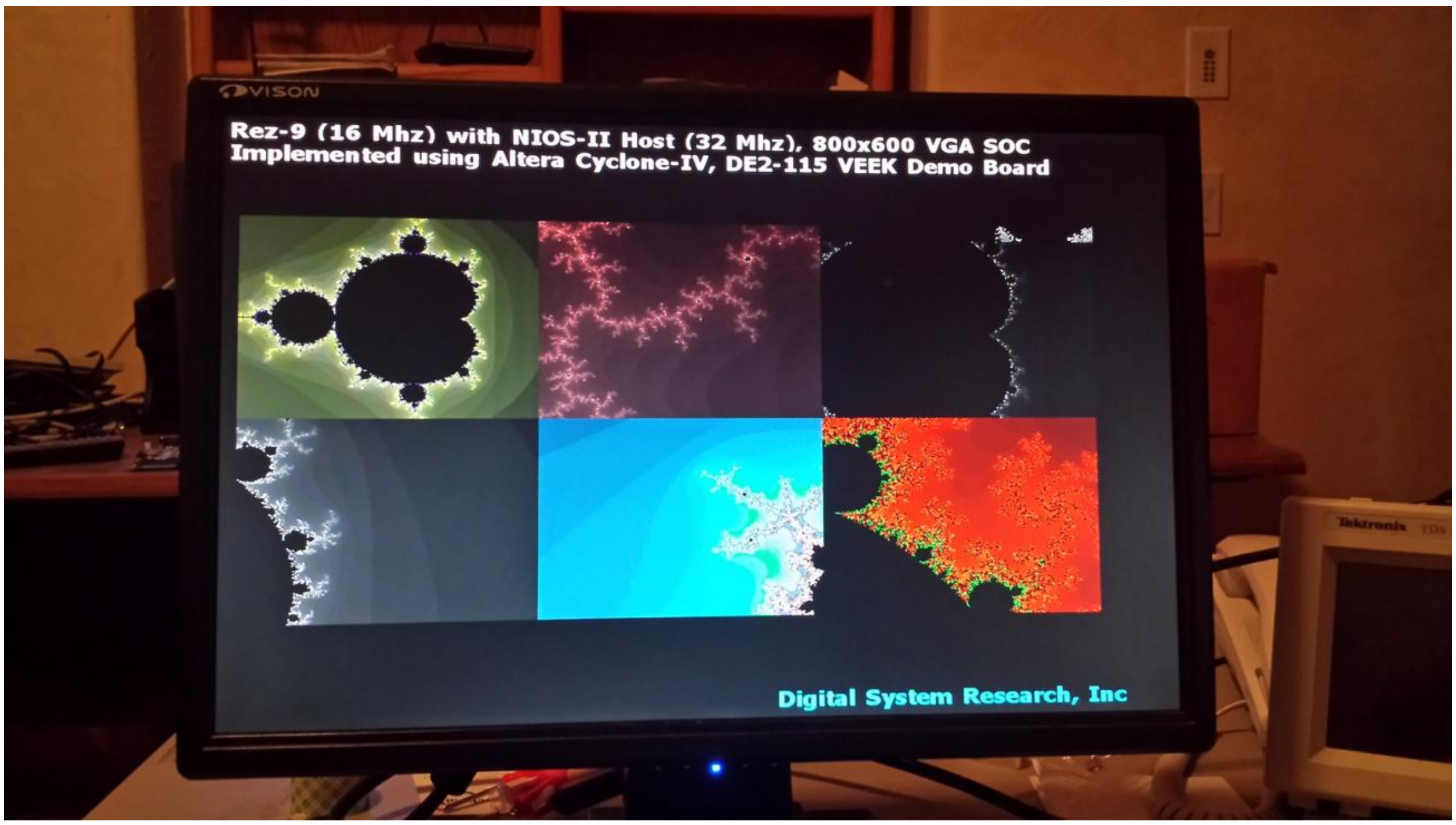

Figure 3. Photo of the Mandelbrot demo created using general purpose residue calculations. The code execution is hybrid. Complex number calculations are performed entirely in residue format using the newly developed fractional residue arithmetic. The threshold comparison is also in residue. The iteration loop count was processed using binary! Mandelbrot fractals require highly precise, iterative arithmetic to generate; the Rez-9/18 exceeds the range of extended precision floating point in this application.

Modern technology turns the tables for the consideration of RNS based processing. When once RNS was considered as a possible solution to construct a single binary multiplier, today's technology offers hundreds, even thousands of binary multipliers now available to resurrect the possibility of RNS processing! The tables have indeed turned.

## A new paradigm for RNS processing

The goal of the Rez-9 ALU is to provide a complete set of arithmetic instructions thereby allowing fully general purpose operation. To that end, it is necessary to implement arbitrary integer division as well as fractional division. For flexibility, a scaling function and the ability to change the fractional resolution are also desirable.

But to what aim? It was discovered that in order to promote the new form of arithmetic, it's necessary to provide researchers a full set of capabilities. In addition, a full set of RNS operations provide a preferred environment for manipulating values. For example, RNS excels at high precision, wide word processing that may exceed a standard binary data type, making the manipulation of these values cumbersome. An eco-system will support a full extent of services, libraries, utilities, and the likes, making it is easier manipulate these RNS values. Finally, the RNS number system is best computed with machines, so machines are needed to accelerate familiarization and experimentation of the new form of arithmetic.

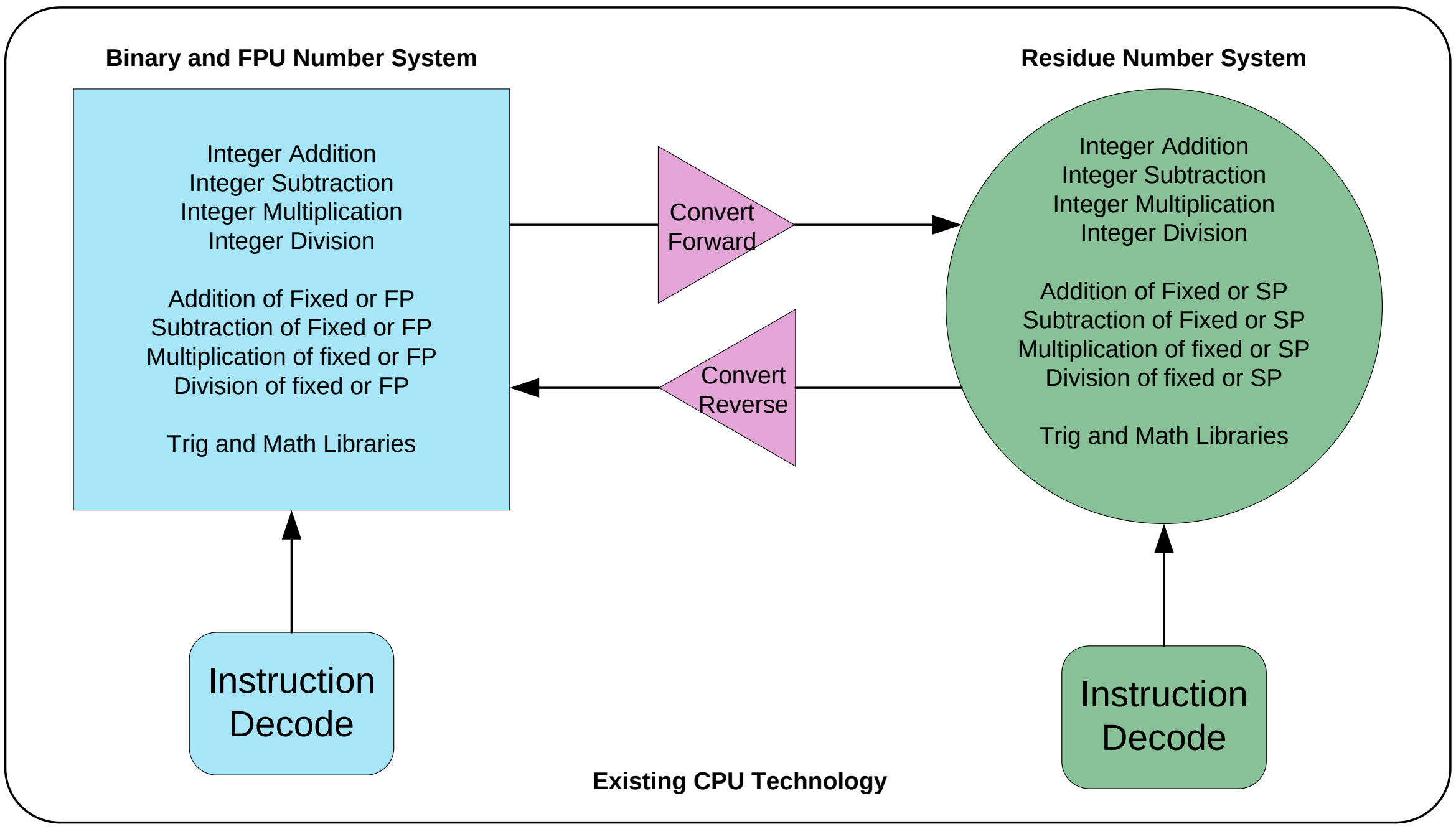

Figure 4

Almost accidentally, and perhaps obviously, a subtle discovery was made in the implementation of the RNS ALU as *co-processor* to a binary CPU, as illustrated in Figure 4. We find that a symbiotic relationship exists between the binary CPU and the RNS ALU. When comparing the strengths and weaknesses of each number system we often see direct tradeoffs. In many cases, when the binary system excels at a specific arithmetic operation, the RNS ALU does not; conversely, when the RNS ALU excels at some operation, the binary CPU does not. This tradeoff allows us to sub-divide a problem to take advantage of each arithmetic domain, thereby speeding execution over binary arithmetic alone.

A new computing paradigm has been invented. Because general purpose residue calculations are now possible, and because direct conversion of fractional values is defined and efficient, an RNS co-processor attached to a standard binary CPU is practical. The new system allows repetitive numeric calculations to be performed in residue format when such calculation is more efficient than binary calculation alone. Because the RNS ALU can provide a full range of operations, this greatly reduces the need for conversion. For many applications, data conversion overhead is negligible, unlike the case of the doomed prior art RNS strategy of the 1960s.

But unless the RNS ALU can be shown to be of significant advantage in certain operations, it too may fail to be implemented in commercial hardware. An RNS system that simply works is not good enough, it must be better in some significant way. Entire ecosystems supporting RNS computation are not likely to emerge without reason. But times change, and so do needs! Remember Google's problem.

## The new "fast" operations in RNS

In this paper, we do not explain RNS arithmetic in detail, but it helps to know the relative speed of basic arithmetic. To be brief, we can speak of a few basic operations. For example, addition in RNS is known to be fast despite word width; the reason is that all digit operations are performed in parallel, at the same time, and without carry. Hence, we refer to this type of operation as a parallel array computation, or PAC for short. In all standard texts covering RNS, it is shown that integer addition, subtraction and multiplication are PAC type operations. In fact, PAC operations are the key feature which motivates researchers to explore RNS.

But what is the speed of fractional multiplication? Fractional multiplication in the Rez-9 supports a fixed point number format. This operation is referred to as a "slow" operation since it takes a multiple number of clock cycles in general. For fractional multiplication, and for machine architectures similar to the Rez-9, a basic rule of thumb is to estimate a number of clocks equal to the number of digits (wide) of the working register. For the Rez-9, this number is 18 clocks, since the basic Rez-9/18 version supports 18 residue digits. This is not particularly fast, however, it is surprisingly competitive with modern CPU's, considering the equivalent binary range of the Rez-9 fixed point representation is roughly equal to extended double floating point numbers (~64 bits)!

So what are the advantages? First, consider that addition and subtraction of RNS fixed point fractions are PAC type operations. That is, these operations only require a single clock cycle regardless of data width. Interestingly, multiplying an integer type to a fractional type is also a PAC operation, known as scaling. Therefore, addition, subtraction and scaling of fixed point fractions are fast and power efficient PAC operations. These are quite nice advantages to have!

But when we consider product summation operations in RNS, we make an important discovery. We find that like the Google TPU, we can trade normalization of every fractional multiplication for a single normalization executed as a last step. This means we eliminate all "slow" normalization operations for every fractional multiplication during a product summation. In fact, we find that product summations comprise a series of PAC operations, where each multiply and each accumulate take only 1 clock regardless of data precision. After accumulation is complete, the final summation is normalized, which is a slow operation, but is easily pipelined. This fact remains true regardless of the number of product terms, and despite the width, or precision, of the data! This is indeed a revolution. The long awaited rewards of processing arithmetic using RNS have become *evident*.

## The case for an RNS based TPU

It is evident we can construct an RNS TPU in a very similar manner as Google's TPU. In the first place, we claim that all required arithmetic operations, with the exception of the activation step, have been clearly defined, developed and tested by DSR. So no missing gaps in the primary arithmetic exist. For the activation step, some of the more simple functions can be supported easily, being most likely integrated into the RNS normalization step. For more complex Sigmoid functions, more research is needed. At DSR our experience has always shown additional research bears fruit! The reason is the new RNS paradigm has blown the doors wide open in terms of arithmetic processing in RNS.

In figure 5, we disclose a basic block diagram for an RNS based TPU derived from the original Google TPU, but with support for multiple layers of circuitry. Each RNS digit is encoded into an 8 bit word, so we may simply re-use the majority of the TPU circuitry for the matrix multiplier and accumulator sections. Each layer represents a distinct RNS digit, and is referred to as a “digit slice”.

Organizing RNS circuits into distinct digit slices is a powerful feature of RNS. As shown in figure 5, arithmetic circuitry for each residue digit modulus is separated in its own layer, and is operated in isolation to all other digits until the step of normalization. This type of data flow is possible because all product summation operations are PAC operations initially. After final accumulation, the residue digits are then brought together into a pipelined normalization and activation unit. Once normalized, digits can be separated again for purpose of storage or re-processing. It is interesting to note each digit of a register can be stored in a separate memory sub system if we desire, as this might be area and space convenient. *If we duplicate the TPU architecture for each digit slice, and each digit slice is operated in parallel, we can claim that speed and efficiency of the Google TPU is preserved in the new RNS TPU design, even while precision of data is significantly increased.*

Several differences are noted, such as the pipelined conversion apparatus sandwiched between the host interface and the RNS TPU shown in purple in Figure 5. Such conversion circuitry may be fully pipelined in this design, to allow full data rates to the DDR3 memory subsystem, or internal register file (unified buffer). It should be noted that conversion pipelines can be implemented using the same 8x8 bit multipliers and 8 bit additions/subtractions (the Rez-9 uses 9x9 bit multipliers and 9 bit addition/subtraction).

Other tradeoffs must be considered. For example, a basic conversion pipeline is composed of the same 8x8 multipliers whose number will grow as the square of the word size (number of digits). For the Rez-9, the word size is 18 digits, providing a total working precision of approximately 62 bits wide. The basic forward pipeline will therefore need around $18^2/2 = 162$ multipliers. Thus, conversion pipelines should not slow, nor impose significant resource issues to the new RNS TPU design. In another RNS design tradeoff, it is typically advantageous to process with a “double” data width, which means data is stored and manipulated in double wide, fully extended registers. This will increase energy compared to storing only a truncated format. This increase is basically a constant factor, as this requirement doubles the number of digit slices that is required for a given binary precision. In the future, more advanced RNS hardware designs will aim to minimize these and other issues. For example, the utilization of dedicated modular multipliers can reduce local data paths by half, which counter acts the need for double width digits, and does so in a manner that maximizes parallelization, and increases speed.

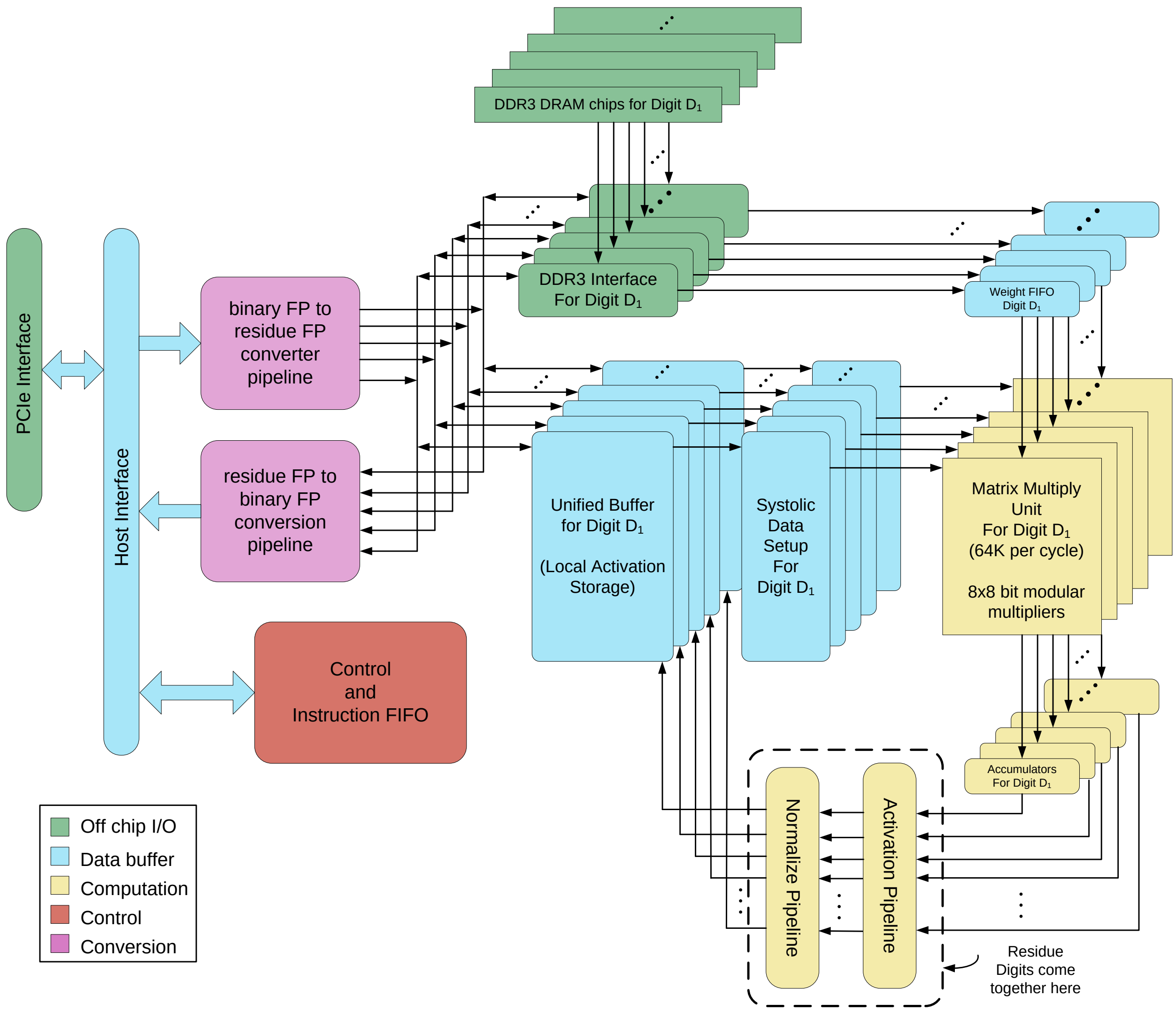

## RNS based TPU Version 1.0

Figure 5.

Illustration shows the proposed RNS TPU capable of Google TPU speeds, yet supporting much wider precision of data. The RNS TPU is nearly identical in architecture except for the inclusion of fractional value conversion pipelines shown in purple, and the addition of fixed MOD functions, either integrated into each 8x8 bit multiply, or inserted as a final step just after accumulation (not shown). The basic tradeoff of each approach is 1) double width buses and increased delays of adding double width integers (same as TPU), versus 2) the delays in implementing the fixed MOD function within each 8x8 bit multiplier and adder. This second approach might be even faster.

Note that residue digits can be separated for most operations, supporting the concept of “digit slices”. Each digit slice is essentially a copy of a Google TPU, without the step of normalization and activation. Since residue digits are encoded in a small data width, say 8 or 9 bits wide, each digit can operate in an identical manner as the Google TPU for most arithmetic operations. When product summation is complete, all residue digits are brought back together for the step of normalization and activation. Afterwards, RNS digits are routed back to their respective digit slice. *Speed and efficiency is preserved, while data precision is increased*™.

## Low power and low area

*"Google would try just about anything if it thought it would help boost compute, network, or storage capacity while also reducing cost and energy consumption"* [5].

Not only is performance a primary factor, but the energy it takes to maintain peak performance is equally important. Several studies have examined the low power operation of modular RNS circuits; each study demonstrates a significant power savings of modular RNS circuits [4][9][15].

Moreover, high efficiency is proven out by the Google TPU itself since the TPU is approximating RNS operation by operating on a very small data width. Looking at figure 5, one can see a linear increase in precision (in terms of equivalent bits) will result in a linear increase in power and circuit area, since increasing precision is simply adding more digit slices. Arithmetic circuits performing binary arithmetic do not exhibit this feature, since increasing data width increases carry propagation and significantly increases area.

## Summary

The performance of the RNS TPU is now evident. The arithmetic circuits have been developed and proven by DSR. The RNS TPU should expect roughly the same speed and efficiency of the Google TPU, yet can scale to perform calculations on much larger data widths. The binary oriented Google TPU cannot scale in this way due to handling carry propagation.

The RNS TPU provides a leap forward in performance and efficiency of *wide precision* neural network accelerators. The new form of RNS arithmetic is very efficient for fractional product summations despite the number of products, and width of the data. Research still remains in some areas of RNS circuit optimization; for example, ASIC libraries which support dedicated modular circuits may provide additional performance and efficiency. Even without this optimization, the RNS TPU can use a Google TPU for each digit slice, and provide for increased precision by operating multiple digit slices in parallel.

Also convenient is the highly specific nature of the RNS TPU application, such that eco systems for general purpose RNS computation are not required. Completely general purpose RNS operation is not required in this application, only fractional product summation and the application of a suitable activation function.

If history is any indication, it's not too late to consider a new number system for applications requiring large amounts of fractional product summation.

## About the author

Eric Olsen is the chief researcher and scientist at DSR. He has been engaged in the study and development of RNS arithmetic and RNS based machine computation for over 25 years. After 20 years of studying the basic RNS system, Eric finally asked the question, "what about RNS fractions?"

## Bibliography & references cited